\documentstyle[preprint,aps]{revtex}
\renewcommand{\baselinestretch}{1.0}
\begin{document}
\draft
\tolerance=10000
\hfuzz=5pt
\title{The Das-Mathur-Okubo sum rule for the charged pion polarizability
in a chiral model}
\author{S. P. Klevansky}
\address{Institut f\"{u}r Theoretische Physik, Universit\"{a}t Heidelberg,\\
Philosophenweg 19, D-69120, Heidelberg, Germany.}
\author{R. H. Lemmer\cite{RHL}}
\address{Institut f\"{u}r Kernphysik, Forschungszentrum J\"{u}lich,\\
P.O. Box 1913, D-52425, J\"{u}lich, Germany.}
\author{ C. A. Wilmot}
\address{Nuclear and Particle Theory Group and Physics Department,\\
University of the Witwatersrand, P.O. WITS 2050, Johannesburg, South Africa.}
\date{\today}
\maketitle

\begin{abstract}
The Das-Mathur-Okubo (DMO) sum rule for the polarizability of charged pions is
evaluated for the Nambu-Jona-Lasinio model Lagrangian
in both its minimal and extended forms.
A comparison is made with the results obtained  using 
the same sum rule
from chiral perturbation theory ($\chi$PT),
 approximate QCD sum rule calculations,
explicit calculations
on the lattice by Wilcox, and using the semi-empirical Kapusta-Shuryak
spectral densities.  The $\chi$PT results from Compton scattering are also
given.  We point to a delicate cancellation between the intrinsic and
recoil contributions to $\alpha_{\pi^\pm}$ in the DMO sum rule approach
that can lead to calculated polarizabilities of either sign.
\end{abstract}

\renewcommand{\baselinestretch}{1.0}
\footnotesize\normalsize

The electric and magnetic polarizability coefficients
$\alpha_\pi$ and $\beta_\pi$ of  pions are fundamental constants of strong interaction physics
and their calculation  is an important testing ground for QCD or effective models
thereof\cite{per75,vautherin89,confine92,lavelle94,burgi96,burgiagain,bajc96,bom92,huf97,dorok97},
since these coefficients can be measured experimentally.  There is however, a reasonable
amount of confusion as to their values, and 
a review of the current status of theory and experiment is given  in \cite{pen95}.
 While $\alpha_{\pi}$  and $\beta_\pi$  can be identified
directly from the soft photon limit of the pion Compton scattering amplitude,
such calculations are extremely tedious, even in lowest
order \cite{burgi96,burgiagain,bajc96,bom92,huf97}.
To obtain a physical picture and common understanding of the values 
obtained by several approaches, it is far simpler to use
the  Das-Mathur-Okubo (DMO) current algebra sum rule \cite{das67}, and we
examine this here.
We also accept the  constraint  $\alpha_\pi+\beta_\pi= 0$, valid for
chiral pions \cite{holcomments90},  and concentrate on  the electrical
polarizability of charged pions. For charged pions, Holstein \cite{holcomments90}
has shown that the DMO sum rule
offers an alternative route
for calculating the intrinsic contribution  to $\alpha_{\pi^\pm}$ by recasting it
as\footnote{The integral
on the right carries an extra 1/2 in \cite{wilcox98} due to the conventional
QCD lattice normalization of the spectral densities used there.}
\begin{eqnarray}
  \alpha^{intr}=
2\sum_{i\neq 0}\frac{|\langle 0|\vec d_z| i\rangle |^2}{E_i-E_0}
=-\frac{\alpha}{m_\pi f_\pi^2}
\int_0^{\infty} \frac{ds}{s^2} [\rho_V(s)-\rho_A(s)]
\label{e:sumrule}
\end{eqnarray}
that together with the center of mass recoil contribution \cite{hufner}
determines $\alpha_{\pi^\pm}$:
\begin{eqnarray}
\alpha_{\pi^\pm}=\frac{\alpha}{3m_\pi}<r_\pi^2>+\alpha^{intr}.
\label{e:def}
\end{eqnarray}
In these formulae, $\rho_{V,A}$ are the vector and axial vector
spectral densities,
 $m_\pi$, $\langle r_\pi^2\rangle$ and $f_\pi$ the pion mass, radius squared and decay
 constant, and $\alpha\approx 1/137$ is the fine structure constant.
 The summation in the defining expression  for $\alpha^{intr}$ in
 Eq.~(\ref{e:sumrule}) runs over all electric dipole
 excitations $|i\rangle$ of energy $E_i$ that are connected to the pionic
 ground state by
 the dipole operator $\vec d$, and the minus sign on the second term
anticipates the fact that $\alpha^{intr}<0$ for pions. Physically, 
this comes about \cite{holcomments90} from an
interplay between  negative energy
intermediate states entering the sum in Eq.~(\ref{e:sumrule}), and  the
 contributions of  disconnected photon and pion decay
 diagrams that have to be  subtracted  out \cite{teren73}.

The DMO sum rule result for $\alpha^{intr}$
is valid to ${\cal O}(p^2)$ in the language of chiral perturbation
theory ($\chi$PT)
 \cite{gasleu85} due to the approximations made in its derivation.
This can be seen as follows.  
 By inserting the
 definitions of the Gasser-Leutwyler coupling constants $\bar l_5,\bar l_6$
 of $\chi$PT \cite{gasleu85},
 \begin{eqnarray}
 <r_\pi^2>=\frac{1}{16\pi^2f^2_\pi}\big(\bar l_6-1\big);\quad
   \int_0^{\infty} \frac{ds}{s^2} [\rho_V(s)-\rho_A(s)]
  =\frac{1}{48\pi^2}\big(\bar l_5-1\big)
   \label{e:radius} 
 \end{eqnarray}
into Eqs.~(\ref{e:sumrule}) and (\ref{e:def}), one  retrieves the
$\chi$PT expression at one loop level
(i.e. also to ${\cal O}(p^2)$) for $\alpha_{\pi^\pm}$ as has also been
extracted  from
Compton scattering \cite{burgi96,holcomments90}
 \begin{eqnarray}
(\alpha_{\pi^\pm})_{\chi PT} =\frac{\alpha}{48\pi^2m_\pi f^2_\pi}\big(\bar l_6-\bar l_5\big).
\label{e:chptalpha}
\end{eqnarray}
This takes the numerical value $ 2.7\times 10^{-4}$ fm$^{3}$ using
experimentally extracted values of the $\bar l_i$, and is a direct prediction 
of QCD.
We may
 thus only make a comparison with $\chi$PT to this order in using the
sum rule.    Note that a higher order calculation of $\alpha_{\pi^\pm}$
has been made via Compton scattering \cite{burgiagain}.    This calculation
involves the computation of over 100 Feynman diagrams, and lowers the final
result from $\alpha_{\pi^\pm} = 2.7 $ to 2.4 $\times 10^{-4}$ fm$^{3}$.

The DMO form for $\alpha^{intr}$ in Eq.~(\ref{e:sumrule}) has the
 particular advantage that this
 contribution to
$\alpha_{\pi^\pm}$  can now be estimated using QCD sum rules as in \cite{lavelle94},
or calculated directly on the lattice as in \cite{wilcox98}.
Here, we also implement  the  DMO calculation of $\alpha^{intr}$, as well as
the recoil contribution  to
$\alpha_{\pi^\pm}$, using a model chiral 
Lagrangian  given below to describe the quark dynamics, and we then compare
this 
 with  QCD sum rule results, calculations on the lattice, chiral perturbation
theory, and also a result obtained 
using the empirical Kapusta-Shuryak \cite{kap94} spectral densities
to evaluate the DMO sum rule. Recent data from the ALEPH collaboration 
\cite{aleph1,aleph2} for the spectral densities is also shown.
We base our discussions on the
 extended Nambu-Jona-Lasinio (ENJL)
  Lagrangian \cite{njl,vog91} which
 explicitly includes  vector and axial vector degrees of freedom:
  \begin{equation}
{\cal L}_{ENJL} = \bar{\psi} \big[{\rm i} {\partial{\mkern -10.mu}{/}}
- \hat{m}\big] \psi + G_{1} \Big[ (\bar{\psi} \psi)^2 +
  (\bar{\psi} {\rm i} \gamma_5 \mbox{\boldmath$\tau$} \psi)^2 \Big]
 - G_{2} \Big[ (\bar{\psi} \gamma_{\mu} \mbox{\boldmath$\tau$} \psi)^2 +
  (\bar{\psi} \gamma_{\mu} \gamma_5 \mbox{\boldmath$\tau$} \psi)^2 \Big].
\label{e:lagra}
\end{equation}
 Here $\bar\psi=(\bar u,\bar d)$ is the $SU(2)$
quark flavor doublet, $\mbox{\boldmath$\tau$}$ Pauli isospin matrices, $G_1,G_2$ coupling constants,
 and $\hat{m}$ a common current quark mass that
explicitly breaks chiral symmetry. Together with a regulating cutoff $\Lambda$,
these parameters are fixed by reproducing the pion mass and  decay constant, quark condensate
density, and the axial coupling constant of the constituent quarks.

The spectral densities  
$\rho_J(s)=-(1/4\pi)Im\tilde\Pi_J(s)$ at four-momentum transfer squared
$s=q^2$ that enter into Eq.~(\ref{e:sumrule})
are
obtained from  the two point polarization functions
of the vector and axial vector currents $J^a_\mu(x)=
V^a_\mu(x)$ or $A^a_\mu(x)$ of isospin index $a$ and have the tensor structure
$\tilde\Pi^J_{\mu\nu;ab}(s)=\tilde\Pi_J(s)(g_{\mu\nu}-q_\mu q_\nu /q^2)\delta_{ab}$.
These amplitudes are purely spin one
(``transverse'') in the chiral limit due to current
conservation, $\partial^{\mu} J_\mu^a(x)=0$. 
Their difference satisfies the  unsubtracted dispersion
relation \cite{wein67,nar89}
\begin{eqnarray}
\frac{1}{4}\big[\tilde\Pi_A(s)-\tilde\Pi_V(s)\Big]
=\int_0^\infty \frac{dt}{t-s-i\epsilon}[\rho_V(t)-\rho_A(t)].
\label{e:dispersion}
\end{eqnarray}
We now specialize to the case where the dynamics of the  $J^a_\mu(x)$
are assumed to be
governed by ${\cal L}_{ENJL}$.
Working to  leading order in $N_c^{-1}$ \cite{quack,dstl95},
 both the lowest order
one-loop expressions $\Pi_{V,A}$, and the  resummed
expressions for the polarization functions $\tilde\Pi_{V,A}$ as given by their
Bethe-Salpeter equation, can be explicitly shown to satisfy  the unsubtracted
dispersion relation, Eq.~(\ref{e:dispersion}), under Pauli-Villars
regularization\footnote{In which case the associated spectral density
difference in Eq.~(\ref{e:dispersion}) mimics \cite{dkl96} the QCD
asymptotic behavior 
$\rho_V(t)-\rho_A(t)\sim {\cal O}(s^{-2})$ \cite{flor79,bra93}.} \cite{dkl96}.
Using this dispersion relation,
  the required integral in Eq.~(\ref{e:sumrule}) can  be rewritten as
  \begin{eqnarray}
\int_0^\infty \frac{dt}{t^2}[\rho_V(t)-\rho_A(t)]=
\frac{1}{4}\frac{\partial}{\partial s}\Big[\tilde\Pi_A(s)-\tilde\Pi_V(s)\Big]_{s\rightarrow 0}.
\label{e:equivalent}
\end{eqnarray}
The difference of polarization amplitudes in the ENJL model is found by direct
 calculation to be given by the relation (Ward identity \cite{brz94})
\begin{eqnarray}
\frac{1}{4}[\tilde\Pi_A(s)-\tilde\Pi_V(s)]=f^2_\pi F_P(s)F_V(s)F_A(s)
\label{e:difference}
\end{eqnarray}
in terms of the pion electromagnetic form factor $F_P(s)$ of the minimal NJL
 model ($G_2=0$) and the vector and
axial vector form factors $F_V(s)$ and $G_A(s)$ of the constituent quark currents.
Closed expressions for the latter two form factors are obtained by summing the
associated Bethe-Salpeter equation to all orders in the coupling $2G_2$. One
finds
\begin{eqnarray}
F_V(s)=[1+2G_2\Pi_V(s)]^{-1} ;\qquad G_A(s)=[1+2G_2\Pi_A(s)]^{-1}=g_AF_A(s)
\label{eq9}
\end{eqnarray}
in terms of  the irreducible one-loop polarization functions $\Pi_{V,A}$. 
Here $g_A=G_A(0)$, or $1-g_A=8G_2f^2_\pi$, is the
axial coupling constant \cite{volkov84,preprint,klim90,ebertvolkov83}. 
The derivative on the right hand side of
Eq.~(\ref{e:equivalent}) that is required to evaluate the integral explicitly is determined by the
behavior of the form factors at vanishingly small momentum transfers.
This can be found in terms of the  $\rho$ and $a_1$ meson  masses $m_{V,A}$ and their
common coupling constant $g_\rho$ to the quarks that can be identified from the
poles and  residues in the corresponding resummed propagators for the
vector and axial vector modes $-iD_{V,A}(s)=\{-2iG_2F_V(s),-2iG_2G_A(s)\}$
after introducing  low-energy
expansions \cite{klim90} for the $irreducible$  polarizations that they contain: $\frac{1}{4}\Pi_V(s)\approx-g^{-2}_\rho
s$ and $\frac{1}{4}\Pi_A(s)\approx-g^{-2}_\rho(s-6m^2)$. Here $m$ is the self-consistent
quark mass of the ENJL model. One finds $m^2_A=6m^2+m^2_V=m^2_V/g_A$ and
$m^2_V=g^2_\rho/8G_2=g^2_\rho f^2_\pi/(1-g_A)$. The latter result follows by
 eliminating $G_2$ in terms of 
 its expression for $g_A$ given above. This last form for $m^2_V$ reduces to the
 KSRF relation \cite{KSRF66} for the choice $g_A=1/2$.   All of these  
 low-energy relations  also follow immediately by working
 with the bosonized version of
 the ENJL Lagrangian from the start \cite{ebertvolkov83}.
 Introducing the same expansions of the irreducible polarizations into the
 denominators of the form factors
 defined above in Eq.(\ref{eq9}), one obtains 
  $F_V(q^2)= 1+q^2/m^2_V+\cdots$ and
$F_A(q^2)= 1+q^2/m^2_A-g_A(1-g_A)q^2/(8\pi^2f^2_\pi)+\cdots$
after also recalling that \cite{tarrach79}
 $F_P(q^2)=1+g_Aq^2/(8\pi^2f^2_\pi)+\cdots$.  
Using this information  the value of the integral in Eq.~(\ref{e:equivalent})
becomes
\begin{eqnarray}
\int_0^\infty \frac{ds}{s^2}[\rho_V(s)-\rho_A(s)]=\frac{g^2_A}{8\pi^2}+
\frac{f^2_\pi}{m^2_V}+\frac{f^2_\pi}{m^2_A},
\label{e:int}
\end{eqnarray}
The pion radius
parameter is found from the expansion of the ENJL pion form factor
$F_\pi(q^2)\neq F_P(q^2)$
to be
\begin{eqnarray}
<r_\pi^2>=\frac{3g_A^2}{4\pi^2f^2_\pi}+\frac{3}{m^2_V}+\frac{3}{m^2_A}
\label{e:rad}
\end{eqnarray}
for $N_c=3$.
Referring back to Eqs.~(\ref{e:sumrule}) and (\ref{e:def}), one sees
that an  exact cancellation is going to occur between the meson mass terms in the
recoil and intrinsic contributions to the electric polarizability.
The final result for the ENJL polarizability 
 as calculated using the DMO sum rule thus reads
\begin{eqnarray}
\alpha_{\pi^\pm}=
\frac{\alpha g^2_A}{8\pi^2m_\pi f^2_\pi}.
\label{e:ENJL}
\end{eqnarray}
 To the order ${\cal O}(N^0_c)$ and ${\cal O}(p^2)$ we are
 working,\footnote{Since  $G_2\sim N_c^{-1}$ in order to
preserve the correct scaling properties of quark and meson masses,
and $f^2_\pi\sim N_c$, one sees that
$1-g_A=8G_2f^2_\pi\sim N_c^0$. The same result follows from $N_c$ 
counting rules \cite{bls93}.}
 this result depends only on the axial coupling constant
 of the constituent quarks besides the pion parameters.
The lesson from this is that
there is a delicate balance between the recoil
and intrinsic contributions,
 so that at a more fundamental level, one should probably  compute both terms to the same level of
approximation, instead of relying, for example, on the known experimental
 pion radius squared for the first term. As in \cite{wilcox98},
  this can give {\it negative} values for $\alpha_{\pi^\pm}$. 
 If, in addition, we insist
that the KSRF relation \cite{KSRF66} should also hold between the meson
parameters  generated by the
ENJL model in its low energy regime, then we have seen that\footnote{ Estimates of $g_A$ using the Adler-Weisberger
sum rule \cite{bls93,peris93}, suggest a much larger ($\sim 0.8-0.9$) value for
$g_A$.}
$g_A=1/2$, and Eq.~(\ref{e:ENJL}) becomes a
 parameter-free
result for the charged pion polarizability, of magnitude $\alpha_{\pi^\pm}\approx
1.5\times 10^{-4}$ fm$^3$ for the physical constants.  This value
is listed in
 Table I. 

 The  expression for $\alpha_{\pi^\pm}$ as obtained from  Compton
  scattering calculations using the NJL model \cite{bajc96,dorok97}
  also reduce to Eq.~(\ref{e:ENJL}) (with $g_A$=1) when evaluated to
  leading order ${\cal O}(p^2)$.  
 This  confirms
   that  the integrals over the model spectral densities behave correctly
  and  is
an important self-consistency check, since these spectral functions
give the
density of states in the $\bar q q$ continuum of the
non-confining ${\cal L}_{ENJL}$, and not that of the physical $\rho\rightarrow \pi\pi$ and
$a_1\rightarrow \pi\rho$ continua. Notwithstanding this unphysical
feature, the sum rules continue to be obeyed. The  reason for this
 is
that the model polarization amplitudes 
 satisfy 
unsubtracted dispersion relations like Eq.~(\ref{e:dispersion}) and the
Ward identity of Eq.~(\ref{e:difference}) under gauge-invariant regularization
schemes (such as  Pauli-Villars or dimensional regularization).
 This in turn allows one to
re-express
the sum rules in terms of the zero momemtum transfer values of the
model polarization amplitudes and their derivatives
 that are not sensitive to 
 the non-confining nature of the model.
 Fig.~\ref{f:densdiff} compares the behavior of the vector minus axial vector
spectral density  difference  
  for the ENJL model with (a)   the
semi-empirical fits of  Kapusta and Shuryak \cite{kap94},
and (b) recent data from the ALEPH collaboration \cite{aleph1,aleph2}.
Both theoretical
 sets of curves satisfy  Weinberg's first sum rule \cite{wein67}.
The ENJL model density difference accomplishes this
  by having considerably smaller and wider  resonant peaks
at the (model) vector masses $m_V=793$ MeV and $m_A=1122$
MeV.   Note, however, that the value that we obtain
 for $\alpha_{\pi^\pm}$ using the
Kapusta-Shuryak densities is {\it negative} as is seen in Table I.
   One sees that the vector peak of the ALEPH data  is well-modelled by
the phenomenological form  of Ref.\cite{kap94}.
    At large $s$, the uncertainties in the data
are too large to make it possible to evaluate the sum rule precisely, for
$s\rightarrow\infty$ \cite{aleph2}.

We next return to the NJL model and 
examine the sub-leading  in $N^{-1}_c$ contributions to
Eqs.~(\ref{e:rad}) and (\ref{e:int}) and show that they are not small.
We illustrate these corrections in the case of  the minimal NJL model
($G_2=0$ therefore $g_A\rightarrow 1$ and $m^2_V,m^2_A\rightarrow \infty$)
that only supports  $\pi$ and $\sigma$ meson modes.
Such a calculation has already been performed for the pion radius to this
order, and one has found \cite{hippe95,lemteg95}  
\begin{eqnarray}
  \langle r^2_\pi\rangle= \frac{3}{4\pi^2f^2_\pi}
-\frac{1}{16\pi^2f^2_\pi}\big(\ln\frac{m^2_\pi}{m^2_\sigma}+\frac{17}{6}\big)
+ O(m_\pi^2), 
 \label{e:newrad}
  \end{eqnarray}
which is necessary for evaluating the recoil contribution to
$\alpha_{\pi^\pm}$.  Note that the $f_\pi$  which enters this expression
 (via a Goldberger-Treiman relation that is itself correct
 \cite{dstl95} to ${\cal O}(1/N_c)$)
 refers to a pion weak decay constant that
is correct to ${\cal O}(1/N_c)$. This comes about because  $\langle r^2_\pi\rangle$
follows from differentiating an NJL electromagnetic form factor  for the
pion that includes ${\cal O}(1/N_c)$ corrections, and therefore
contains a $\pi qq$ coupling constant $g_{\pi qq}$ that is also correct to this order. Otherwise
the modified form factor does not normalise properly to unity at $q^2=0$. This is demonstrated
explicitly in \cite{lemteg95}.

A similar analysis can be made for the spectral
density function integral.   This is briefly sketched here.
Corrections arise from two types of diagrams \cite{dstl95}: (i) single meson loop  dressings
of the irreducible one-loop quark polarization diagram,
and (ii) meson pairs exchanged
between $\rho $ and $a_1$ vertices.
While type (i) is essential for maintaining the transverse
nature of the vector polarization diagrams to ${\cal O}(N^0_c)$, the subset (ii)
of irreducible two meson loop diagrams  is singled out by the fact
that their derivatives at $q^2=0$ contain chiral logarithms \cite{mp78}
that diverge  as $m^2_\pi\rightarrow 0$  due to the
pion pole in the  propagator $D_\pi$ of the pion.
So they dominate
over the derivatives\footnote{In fact the
type (i) diagrams are strictly constants i.e. are independent of the external meson momentum
in the bosonized version of the NJL model and thus give no contribution at all to
the difference of polarization derivatives in Eq.~(\ref{e:derivativediff})
in that case.}
of type (i) in the chiral limit.
Denoting the meson pair exchange diagrams with the
 coupling to the external channel factored out by
  $-i\Pi^{\rho\pi\pi}_{\mu\nu}(q^2)$ and
  $-i\Pi^{a_1\pi\sigma}_{\mu\nu}(q^2)$ respectively, one has 
   \begin{eqnarray}
 (-\frac{i}{2}g_\rho)^2  \big[\frac{1}{i}\Pi^{\rho\pi\pi}_{\mu\nu}(q^2)\big]=\int\frac{d^4l}{(2\pi)^4}
   \Gamma^{\rho\pi\pi}_{\mu}(l,l+q) D_\pi((l+q)^2)D_\pi(l^2)
   \Gamma^{\rho\pi\pi}_{\nu}(l+q,l)
 \label{e:mesonloop}
 \end{eqnarray}
for the $\rho$ channel.  A similar expression holds for the $a_1$ channel.
  The leading  contribution in chiral logarithms to this
  is found most directly
 by using  the
 bosonized version  \cite{ebertvolkov83,volkov84} of the
 Lagrangian (\ref{e:lagra}) to identify the
 effective meson vertices
 $\Gamma^{\rho\pi\pi}_\mu$ and $\Gamma^{a_1\pi\sigma}_{\mu 5}$.
  These vertices are generated by the Lagrangians  
 \begin{eqnarray}
 \delta{\cal L}_{\rho\pi\pi}=-g_\rho \epsilon_{abc}\pi_a\partial_\mu\pi_b\rho^\mu_c\;;\quad
 \delta{\cal L}_{a_1\pi\sigma}=g_\rho\sigma(2\vec a_1^\mu\cdot\partial_\mu\vec\pi
  +\vec\pi\cdot\partial_\mu\vec a^\mu)
  \end{eqnarray}
  where  ($\sigma, \vec \pi,  \vec\rho, \vec a_1$)
  are  isoscalar and isovector meson fields
  and  $g_\rho$ is the common vector meson
 coupling constant.  Filling in the details, one finds the convergent contribution
 \begin{eqnarray}
 \frac{1}{4} \frac{\partial}{\partial s}[\Pi^{a_1\pi\sigma}_T(s)&-&\Pi^{\rho\pi\pi}_T(s)
 ]_{s\rightarrow 0} = \frac{1}{8\pi^2}
 \int_0^1 d\alpha\alpha(1-\alpha)\ln\Big\{\frac{m^2_\pi(1-\alpha)+m^2_\sigma\alpha}{m^2_\pi}\Big\}
 \nonumber\\
 &=&- \frac{1}{48\pi^2}
 \Big(\ln\frac{m^2_\pi}{m^2_\sigma}
 +\frac{5}{6}\Big)+O(m^2_\pi),
 \label{e:derivativediff}
 \end{eqnarray}
 to be added to Eq.~(\ref{e:equivalent}), after projecting out onto the
 transverse parts of the meson loop polarization diagrams.
  Here $ m_\pi$ and $m_\sigma=(m^2_\pi+ 4m^2)^\frac{1}{2}$
  are the $\pi$ and $\sigma$ meson masses of the bosonized NJL model \cite{ebertvolkov83}.
 The revised value of the integral (\ref{e:int}) for the NJL case ($g_A=1$)
  now reads 
 \begin{eqnarray}
 \int_0^\infty\frac{ds}{s^2}[\rho^V(s)-\rho^A(s)]_{(\rm quark + meson\; loops)}
 &=&
 \frac{1}{8\pi^2}-\frac{1}{48\pi^2}
 \Big(\ln\frac{m^2_\pi}{m^2_\sigma}
 +\frac{5}{6}\Big)+O(m^2_\pi).
 \label{e:newint}
 \end{eqnarray}
  In Eqs.~(\ref{e:newint}) and (\ref{e:newrad}) we  have also retained
  chiral ${\cal O}(N^0_c)$ corrections, which are pure numbers in addition to
  the chiral  logarithms that arise from the meson loop, in order 
to be completely
  consistent in $N_c$ power counting.  From Eqs.~(\ref{e:radius}),
  one obtains values for the two  scale-independent
 Gasser-Leutwyler coupling constants 
 $\bar l_5+\ln(m^2_\pi/\mu^2)=7-5/6=37/6$
 and $\bar l_6+\ln(m^2_\pi/\mu^2)=13-17/6=61/6$,
   (here evaluated at the natural scale
 $\mu=m_\sigma \approx 500-600$ MeV of the NJL model)
 that are in the  form
 given by
   chiral  perturbation theory \cite{gasleu85}.
  The chiral logarithms cancel in the revised
 value of their  difference   which
  now becomes $\bar l_6-\bar l_5=4$ in Eq.~(\ref{e:chptalpha}), down from
$13-7=6$, due to the remaining ${\cal O}(N^0_c)$  mass-independent
corrections that
 the $\bar l$'s now contain. This is to be compared with the empirical
 difference of $16.5-13.9=2.6$.
 The ${\cal}O(N^{-1}_c)$ meson loop corrections are thus
 large and {\it negative}
  in this case, leading to a considerable {\it decrease}
 in the predicted NJL value  for $\alpha_{\pi^\pm}$, see Table I.

To summarize, the charged pion polarizability is examined 
in the NJL and ENJL models
 and compared with the results obtained in $\chi$PT, 
which serves as a benchmark of QCD.   For completeness, the Compton
scattering result for this quantity at two loop order is also quoted.
hile this does not suffer from any obvious cancellation problems, it 
involves the computation of more than 100 graphs. The
NJL model can be directly be compared with $\chi$PT to one loop order,
 in that the formal
expression (4) holds in the NJL model with low energy constants evaluated within the model.
This, in turn, leads to a much higher value of $\alpha_{\pi^\pm}
$ than does 
$\chi$PT, unless one includes the next order meson loops in the NJL model
$N_c^{-1}$ expansion.   In the ENJL model, to lowest order, the final
result is dependent on the value of $g_A$.    In an analysis according to the
division into intrinsic and recoil contributions,
    one notes that the intrinsic contribution to the polarizability
   changes by about a   factor 3 from the
  minimal  NJL calculation ($G_2=0$) to using the empirical densities of Kapusta and Shuryak.
  The negative results for $\alpha_{\pi^\pm}$ in the latter case, as well as for
  the lattice calculations, reflect the delicate balance
   between the
  two relatively large recoil and intrinsic contributions
  to  $\alpha_{\pi^\pm}$ that are of opposite sign.
   This   emphasizes the need for calculating $both$ terms in the sum rule
   approach to   the same level of approximation.
The ENJL result ($-11.0$) for $\alpha^{intr}$ is itself nicely bracketed by the QCD sum rule
result ($-9.6$) and chiral perturbation theory ($-12.6$),
all in units of $10^{-4}$ fm$^3$, having been considerably increased over the
minimal NJL case ($-5.9$) by  the presence of 
 the vector and axial vector modes.
 We have set $g_A=1/2$ in the ENJL case in order to comply with the KSRF
 relation.  However, even allowing $g_A$ to fall through its full range
$1\ge g_A\ge 0$ (while scaling $m$ like $g_A^{-1/2}$ to keep the same
$f^2_\pi$) gives values of the integral that only range between 12.7 and 27.5.
In the latter instance the intrinsic and recoil contributions would simply cancel,
leaving a polarizability coefficient of zero.

One of us (RHL) would like to thank Prof. J. Speth, and the  members of the Theory Division at the
Institut f\"ur Kernphysik at the Forschungszentrum, J\"ulich, for their kind hospitality.
Financial support from the Alexander von Humboldt Foundation is also gratefully
acknowledged. This work has also been partially supported by the Deutsche
Forschungsgemeinschaft (DFG) under contract  number HU 233/4-4 and by the German
Ministry for Education and Research (BMBF) under contract number 06 HD 742.


\begin{figure}
\caption[]{The vector minus axial vector spectral densities  as calculated from the ENJL
model (solid curve), and  from the semi-empirical Kapusta-Shuryak expressions of
Ref.~\cite{kap94} (broken curve) versus momentum transfer squared.
The dotted points with error bars are recent experimental values taken from
the ALEPH collaboration.
 The Pauli-Villars scheme
has been used to regulate the ENJL polarization amplitudes.  This
introduces two further artificial thresholds (only the first one is shown)
at  equally spaced intervals
 $4\Lambda^2$   beyond the unphysical
$\bar qq$ threshold that lies at $4m^2=0.28$ GeV$^2$
 for the particular ENJL parameter choice  
$G_1=2.47$ GeV$^{-2}$,  $G_2=3.61$ GeV$^{-2}$,
and a regulating cutoff of $\Lambda= 1.06$ GeV, used in this figure.}
\label{f:densdiff}
\end{figure}                

\begin{table}
\caption[]{Breakdown of intrinsic and finite size contributions to the  
polarizability of charged pions in units of $10^{-4}$ fm$^3$ (1 fm $\approx
197.2$ MeV$^{-1}$) as
obtained from various theoretical approaches,  together with the available
experimental data.
  The original Serpukov II and Mark II data
 analyses quote 
   $(\alpha_{\pi^\pm}\pm\beta_{\pi^\pm})$ without assuming the
  constraint $\alpha_{\pi^\pm}+\beta_{\pi^\pm}=0$. We have simply averaged the
  statistical and systematic error bounds when extracting $\alpha_{\pi^\pm}$. Input: $m_\pi=139.6$
MeV, $f_\pi=93$ MeV, and a typical quark mass $m=324$ MeV from solving the
gap equation \cite {njl} for the ENJL case,
giving $m_V=793$ MeV and $m_A=1122$ MeV; empirical values $\bar l_5=13.9\pm 1.3$, $\bar l_6=
16.5\pm 1.1$  for the $\chi$PT case \cite{gasleu85}, and
the measured value $<r^2_\pi>=0.439$ fm$^2$ \cite{amendolia86} except where  derived
 from a model.}
\label{table of comparison}

\begin{tabular}{lcccc}
Approach&$\rm{Spectral\; density\; integral}$&$\alpha^{intr}
$&$\alpha \langle r^2_\pi\rangle/3m_\pi$&$\alpha_{\pi^\pm}$\\
 &$(\times 10^{-3})$& & & \\
\tableline
$\chi$PT  (Compton scattering to two loops) \cite{burgiagain}
& & & & 2.4 $\pm$ 0.5\\
$\chi$PT$\;[\rm{Eqs.}~(\ref{e:radius},\ref{e:chptalpha})$]&27.2&$-12.6
$&15.2&2.7$\pm$0.4\\
NJL$\;[{\rm{Eqs}.~(\ref{e:int},\ref{e:rad})},\;g_A=1$]&12.7&$-5.9$&11.7&5.8\\
NJL+$O(N^{-1}_c)\;[\rm{Eqs.~(\ref{e:newint},\ref{e:newrad})}$]&15.9&$-7.4$&11.3&
3.9\\
QCD sum rule$\;\cite{lavelle94}$&21.0&$-9.6$&15.2&5.6$\pm$0.5\\
ENJL$\;[{\rm{Eqs}.~(\ref{e:int},\ref{e:rad})},\;g_A=0.5$ (from KSRF)]&23.8&$-11.0$&12.5&1.5\\
Lattice calc.$\;\cite{wilcox98}$&36.3&$-17.1$&15.2&$-2.0\pm1.8\pm1.6$\\
Semi-empirical densities$\cite{kap94}$ &39.4&$-18.2$&15.2&$-3.0$\\            
Lebedev\cite{czech}&-&-&-&$20\pm 12$\\
Serpukov I\cite{antipov1}&-&-&-&$6.8\pm 1.4$\\
Serpukov II\cite{antipov2} &-&-&-&$8.5\pm 4.8\pm 3.5$\\
Mark II\cite{mark90}&-&-&-&$2.5\pm 0.5\pm 0.02$\\
$\gamma\gamma\rightarrow\pi\pi$ measurement\cite{babusci}&-&-&-&$2.2\pm 1.6$\\
\end{tabular}                         
\end{table}

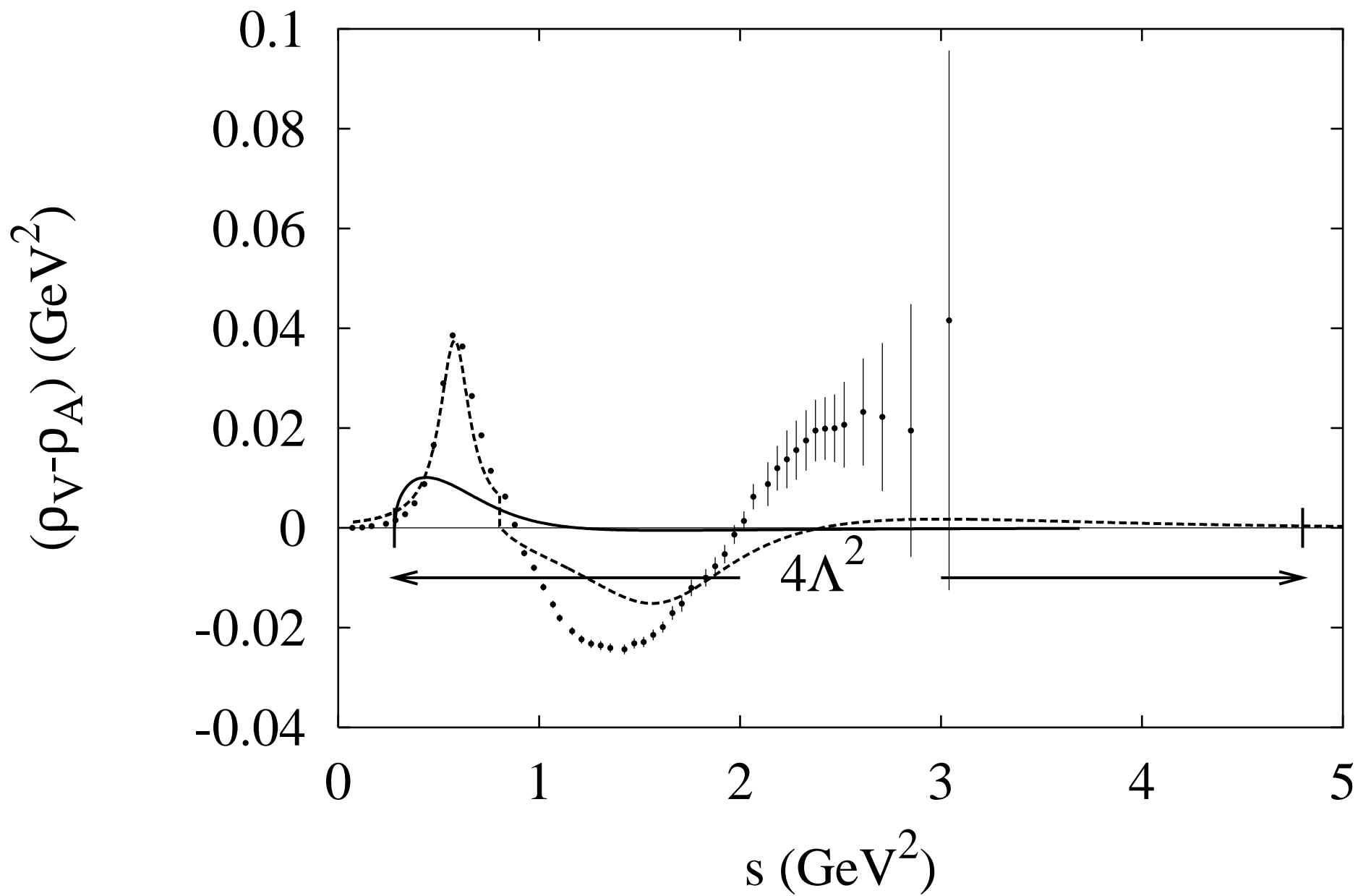
\end{document}